\documentclass[aps,showpacs]{revtex4}
 \def\ket{\rangle}
\def\<{\langle}
\def\>{\rangle}
\begin{document}
\title{Efficient Multi-Party Quantum Secret Sharing Schemes}
\author{Li Xiao$^{1,2}$,
Gui Lu Long$^{1,2,3,4}$\footnote{Corresponding
author:gllong@tsinghua.edu.cn}, Fu-Guo Deng$^{1,2}$, Jian-Wei
Pan$^{5}$}
\address{$^{1}$Department of Physics, Tsinghua University, Beijing
100084, China\\
$^2$ Key Laboratory for Quantum Information and Measurements,
MOE, Beijing 100084, P R China\\
$^3$ Center of Atomic and Molecular NanoSciences, Tsinghua
University, Beijing 100084, China\\
$^4$ Center for Quantum Information, Tsinghua University, Beijing
100084, China\\
 $^5$ Institute for Experimental
Physics University of Vienna, Boltzmanngasse 5, Vienna 9, Austria\
}
\date{\today}

\begin{abstract}
In this work, we generalize the quantum secret sharing scheme of
Hillary, Bu\v{z}ek and Berthiaume[Phys. Rev. A59, 1829(1999)] into
arbitrary multi-parties. Explicit expressions for the shared
secret bit is given. It is shown that in the
Hillery-Bu\v{z}ek-Berthiaume  quantum secret sharing scheme the
secret information is shared in the parity of binary strings
formed by the measured outcomes of the participants. In addition,
we have increased the efficiency of the quantum secret sharing
scheme by generalizing two techniques from quantum key
distribution. The favored-measuring-basis Quantum secret sharing
scheme is developed from  the Lo-Chau-Ardehali technique[H. K. Lo,
H. F. Chau and M. Ardehali, quant-ph/0011056] where all the
participants choose their measuring-basis asymmetrically, and the
measuring-basis-encrypted Quantum secret sharing scheme is
developed from the Hwang-Koh-Han technique [W. Y. Hwang, I. G. Koh
and Y. D. Han, Phys. Lett. A244, 489 (1998)] where all
participants choose their measuring-basis according to a control
key. Both schemes are asymptotically 100\% in efficiency, hence
nearly all the GHZ-states in a quantum secret sharing process are
used to generate shared secret information.
\end{abstract}
 \maketitle

\section{Introduction}
\label{s1}

 The combination of quantum mechanics with information
has produced many interesting and important developments. Quantum
cryptography is one important application. Quantum key
distribution(QKD) concerns the distribution of one-time-pad keys
between distant two parties\cite{bb84}.  With quantum mechanics,
other cryptographic task can be realized. Suppose Alice wants two
parties Bob and Charlie who are at distant places to fulfil
certain task. Alice knows that one of them may be dishonest, but
she does not know who this dishonest guy is. To complete this
task, classical cryptography uses the secret sharing
technique\cite{rcss,rcss2}.  In quantum information, this task can
be achieved by quantum secret sharing(QSS), and it is a fruitful
area of research. Many researches have been carried
out\cite{hbb,gottesman,cleve,karl,band}. It has also been
demonstrated in experiment recently\cite{tittel}. With quantum
mechanics, one can share both  classical  information and quantum
information. In this paper, we consider the issue of sharing of
classical secret information. We specifically consider the QSS
scheme proposed by Hillery, Bu\v{z}ek and Berthiaume(Hereafter we
refer to HBB protocol) \cite{hbb}. In Ref.\cite{hbb}, secret
sharing with 3 and 4 parties have been studied. In the HBB QSS
scheme, the secret sharing is accomplished by using
GHZ-state\cite{ghz}. In this scheme, Alice, Bob and Charlie need
to choose randomly one measuring-basis from  either the
$\sigma_x$-measuring-basis or the $\sigma_y$-measuring-basis
respectively, similar to the BB84 QKD scheme\cite{bb84}. In half
of the cases, nobody chooses the $\sigma_y$-axis or two parties
choose the $\sigma_y$-axis, the measuring results of the three
parties are correlated. In these cases, Bob and Charlie can
combine their measuring-basis information and their measurement
outcomes to determine the results of Alice's measurement. In this
case, Alice's measurement result is used as the secret information
that she wants  Bob and Charlie to share. Generally in order to
establish the secret sharing scheme, a detailed table needs to be
constructed to list all the possible combinations of the
measuring-basis and the possible outcomes of all parties. When the
number of participating parties is large, the construction of such
a table is very tedious and it is also inconvenient to use. In
this paper, we reformulate the HBB scheme in simple mathematical
terms, and the shared secret information becomes  the parity of a
binary string formed by the measured outcomes of the participating
parties. From this formulation, the rules in the HBB QSS scheme
are obtained. For instance for a round of communication to be a
valid one, the number of parties choosing the $\sigma_y$-basis has
to be even, because when even number of parties choose the
$\sigma_y$ basis, the bit value of Alice has a one-to-one
correspondence with the parity of the binary number formed by the
measurement outcomes of the participating parties. From this
mathematical formalism, we generalize the HBB scheme into
arbitrary number parties cases. This  is given in section
\ref{s2}.

In the HBB QSS scheme,  only half of the GHZ-states can be used
for secret sharing. This is the intrinsic limitation of this
scheme. This is similar to the BB84 QKD scheme where only half of
the photons transmitted can be used to generate useful keys. In
QKD, Lo, Chau and Ardehali \cite{chaulo} have proposed a scheme
that increases the intrinsic efficiency to 100\% asymptotically.
In their scheme, they choose one preferred measuring-basis most of
the time, and choose the other measuring-basis, the unfavored
measuring-basis, with a small probability. The events that Alice
and Bob choose the unfavored measuring-basis
 are used later for eavesdropping
checking. This has greatly increased the intrinsic efficiency of
the BB84 scheme. In the limiting case of large numbers, the
efficiency approaches 100\%. This has been shown to be
unconditionally secure\cite{chaulo}. Hwang, Koh and Han\cite{hkh}
have proposed another modification to the BB84 QKD scheme that
also increases its efficiency  to nearly 100\% by letting Alice
and Bob to choose identical measuring-basis according to a common
secret key. This control key is used repeatedly during a QKD
transmission session. Controlled keys have also been used in QKD
in the controlled-order-rearrangement-encryption
scheme\cite{denglong}, where Alice takes one particles from each
Einstein-Podolsky-Rosen(EPR) pair from a group of EPR pairs and
mixes up their orders and sends them to Bob, and Bob recovers the
orders of the particles to get the correct particle correlation of
EPR pairs. Alice and Bob synchronize their action by using a
control key repeatedly. The following example explains the Koh and
Han technique: if the control key is 0101001110, then Alice and
Bob choose their measuring-basis in the following sequence
"$+\times+\times++\times\times\times+$" where "$+$" represents the
vertical-horizontal measuring basis, and "$\times$" represents the
diagonal-antidiagonal measuring-basis. The control key is usually
quite short, say 1000 bits long. It is repeated again and again
until the QKD process ends. In this way, Alice and Bob can always
choose the same measuring-basis. The quantum mechanical nature of
the single photons renders eavesdropping detectable, and the
random nature of the measured results keeps the information on the
control key  safe. It has been shown that the scheme is
secure\cite{hkhproof} for ideal single photon sources. The
essential ingredient in these improvements is to allow the two
parties to use identical measuring-basis as much as possible
Generalizing these two techniques, we have proposed two efficient
QSS schemes that is asymptotically 100\% in efficiency. This is
given in section \ref{s3}. A summary is given in section \ref{s4}.

\section{Multi-Party HBB Quantum Secret Sharing Scheme}
\label{s2}
 We present the $n$-party  HBB QSS scheme first. Suppose there are $n$-parties
taking part in the secret sharing process. It is done by using a
sequence of GHZ-multiplets
\begin{eqnarray}
|\psi\ket_{\rm GHZ}={1\over
\sqrt{2}}(|000..0\ket+|111...1\ket),\label{ghz3}
\end{eqnarray}
where state $|0\ket=|z+\ket$ and $|1\ket=|z-\ket$ are eigen-states
of the spin-projection in the $z$-direction, $\sigma_z$. Alice
keeps one particle and sends the other two particles to Bob and
Charlie each. Then Alice, Bob, Charlie randomly choose from the
$\sigma_x$ and the $\sigma_y$ basis to measure their particles
respetively. The eigen-states of the $\sigma_x$ and the $\sigma_y$
operators are
\begin{eqnarray}
|0\ket_x&=|+x\ket={1\over\sqrt{2}}(|0\ket+|1\ket),
&|1\ket_x=|-x\ket={1\over\sqrt{2}}(|0\ket-|1\ket),\\
|0\ket_y&=|+y\ket={1\over\sqrt{2}}(|0\ket+i|1\ket),
&|1\ket_y=|-y\ket={1\over\sqrt{2}}(|0\ket-i|1\ket).
\end{eqnarray}
Inversely we have the expansion of the eigen-states of the
$\sigma_z$ operator in terms of the $\sigma_x$ and the $\sigma_y$
eigen-basis as follows
\begin{eqnarray}
|0\ket={1\over \sqrt{2}}(|0\ket_x+|1\ket_x), & |1\ket={1\over
\sqrt{2}}(|0\ket_x-|1\ket_x),\label{xbase}\\
|0\ket={1\over \sqrt{2}}(|0\ket_y+|1\ket_y), & |1\ket=-{i\over
\sqrt{2}}(|0\ket_y-|1\ket_y)\label{ybase}.
\end{eqnarray}
We make the convention that the positive polarized states along
$x$- or $y$-axis, are taken as 0, and those along the negative
direction are denoted as 1.  In the HBB scheme, only half of the
GHZ particles can be used for secret sharing. The choice of the
measuring basis plays an important role in judging whether a round
of measurement can be used for secret sharing.

We use a sequence $[b_1(j),b_2(j),...,b_i(j),...,b_n(j)]$ to
denote the measuring basis information for Alice, Bob ..., for the
$j$-th GHZ-state. The number in the bracket, $j$, refers to the
$j$-th GHZ-state in a sequence of secret sharing operations. The
subscript refers to the order number of particles, 1 represents
Alice's particle, 2 refers to Bob's particle and so on. If
$b_i(j)=0$, then the $i$-th party uses the $x$-basis, and
$b_i(j)=1$ means that the $i-$th party uses the $y$-axis. To
obtain the measuring result in such a case, we need to expand the
GHZ-state in the eigen-basis of $[b_i(j), i=1,...,n]$. Using
equations (\ref{xbase}) and (\ref{ybase}),  the $|00\cdots 0\ket$
component can be written as
\begin{eqnarray}
|00\cdots 0\ket=\prod_{i=1}^n\left(\sqrt{1\over
2}(|0\ket_{b_i}+|1\ket_{b_i})\right), \end{eqnarray} and
$|11\cdots 1\ket$ component can be written as
\begin{eqnarray}
|11\cdots 1\ket=\prod_{i=1}^n\left({-i\over
\sqrt{2}}(|0\ket_{b_i}-|1\ket_{b_i})\right). \end{eqnarray} When
the  $y$-basis are chosen by an odd number of participants, the
expansion for $|11\cdots1\ket$  has the following form
\begin{eqnarray}
|11\cdots 1\ket=\pm {i\over
(\sqrt{2})^n}\prod_{i=1}^n\left((|0\ket_{b_i}-|1\ket_{b_i})\right),
\end{eqnarray}
where $+$-sign is for $n=2k+1$ and $-$-sign is for $n=4k+1$ where
$k$ is a positive integer.

Hence the GHZ-state in (\ref{ghz3}) can be rewritten as
\begin{eqnarray}
|\psi\ket_{\rm GHZ}={1\over
2^{(n+1)/2}}\left(\prod_{i=1}^n(|0\ket_{b_i}+|1\ket_{b_i}) \pm
i\prod_{i=1}^n(|0\ket_{b_i}-|1\ket_{b_i})\right), \label{ghzodd}
\end{eqnarray}
for odd number of participants choosing the $y$-basis.
 There is no
cancellation between the 1st product term and the second product
term in Eq.(\ref{ghzodd}), and terms such as $|0i_2i_3\cdots
i_{n-1}\ket_{b_1b_2\cdots b_n}$ and $|1i_2i_3\cdots
i_{n-1}\ket_{b_1b_2\cdots b_n}$ both present in the expansion in
(\ref{ghzodd}). In other words, for a set of measured values
$i_2$, $\cdots$, $ i_n$ measured in measuring-basis $b_2$,
$\cdots$, $b_n$ by the participants Bob, Charlie and so on,
Alice's measured result still has two possibilities. Even if the
$n-1$ secret sharing parties get together and disclose their
measuring-basis information and their measuring outcomes, they
still can not obtain the result of Alice's measurement. For
instance in a 3-party QSS, if Alice chooses the $y$-axis, Bob and
Charlie choose the $x$-axis, the expansion in the eigen-basis of
$\sigma_y$, $\sigma_x$ and $\sigma_x$ (for Alice, Bob and Charlie
respectively) will be
 \begin{eqnarray}
 {(1-i)\over 4}(|000\ket+|011\ket +|101\ket +|110\ket)
                + {(1+i)\over 4}(|001\ket+|010\ket+|100\ket+|111\ket)
                 \end{eqnarray}
 where the basis-orders $y-x-x$ in the subscript are omitted for brevity.
 When Bob and Charlie have definite measured results,
 Alice's result still has two possibilities. For instance, when
 Bob and Charlie's results are 0 and 0 respectively ( along the positive-x axis), Alice's
 result may be 0 (along the positive-y axis) from component $|000\ket$, or $1$ (along the negative-y axis)
  from component
 $|100\ket$. Thus it is not useful for quantum secret sharing.

 When the number of parties choosing the $y$-basis is even,
 we have
\begin{eqnarray}
|\psi\ket_{\rm GHZ}={1\over
2^{(n+1)/2}}\left(\prod_{i=1}^n(|0\ket_{b_i}+|1\ket_{b_i}) \pm
\prod_{i=1}^n(|0\ket_{b_i}-|1\ket_{b_i})\right). \label{ghzeven}
\end{eqnarray}
Because some terms in the second product term in the expansion
have negative sign, they cancel with relevant terms in the first
product term, hence there are only $2^{n-1}$ terms left in
Eq.(\ref{ghzeven}). Among the $2^{n-1}$ terms, the value of the
first bit, which is the result of Alice's measurement, is uniquely
determined by the remaining $n-1$ bit values. In this case when
the $n-1$ parties get together and reveal their measuring-basis
information and the measured results, they can uniquely determine
the bit value of Alice. Unless all $n-1$ participating parties
present, the determination of Alice's bit value is impossible. For
instance, if only $n-2$ parties are present, from their measured
values and measuring-basis information, they can only narrow the
state down to two possibilities in which Alice can have either 0
or 1. Thus it is only when all $n-1$ parties work collectively
that they can get the bit value of Alice.

Summarizing the above observation, the general rule for
multi-party secret sharing are:

1) The number of parties using the $\sigma_y$-basis has to be
even;

2) When the number of parties using $y$-basis is equal to
$2(2k+1)$ where $k$ is a nonnegative integer, the bit value of
Alice is simply the modulo 2 sums of the $n-1$ parties' bit value
plus 1:
\begin{eqnarray}
i_{\rm Alice}=i_1=i_2\oplus i_3\oplus\cdots\oplus i_n\oplus
1.\label{rule1}
\end{eqnarray}
For instance in a 3-party QSS with two parties using $y$ axis,
there is a component $|100\ket$ in the expansion (\ref{ghzeven}) ,
the modulo 2 sum of Bob and Charlie gives 0, then adding 1 gives
1;

3)When the number of parties taking the $y$-basis is $4k$, then
the bit value of Alice is simply the modulo 2 sum of the $n-1$
parties' bit values.
\begin{eqnarray}
i_{\rm Alice}=i_1=i_2\oplus i_3\oplus\cdots\oplus
i_n.\label{rule2}
\end{eqnarray}

Hence the $n$-party HBB QSS scheme can be given as follows: (1)
Alice prepares an $n$-particle GHZ-state (\ref{ghz3}). (2) Alice
keeps one particle at her own hand and sends the rest of particles
to the $n-1$ participants, each party a particle; (3) Each party
chooses randomly from the $x$ or the $y$ measuring-basis to
measure his/her particle. He/she keeps the measured result  and
the measuring-basis information for his/her particle. If the
measured result is up(down) along the measuring-basis, records the
result as 0(1); (4) the above procedures (1)-(3) are repeated many
times until sufficient number of measured results are produced.
This should be at least twice as much as the number of desired
shared bits; (5) after procedure (4), each participants sends the
measuring-basis information to Alice through a classical channel
and upon receiving all the measuring-basis information, Alice
counts the number of parties choosing the $y$-basis. Alice
publicly announces the nature of this number for each round: odd,
or an even number with the form of $2(2k+1)$, or an even number
with the form of $4k$. The exact number of $k$ needs not be
disclosed. If the number is odd, then that round of measurement
result is dropped, and if the number is even, all the participants
keep their measured values and the measuring-basis information for
these events. (6) Alice selects a sufficiently large subsets of
events and asks the participants to disclose their measured values
for these events. From this information, Alice can check if there
exists eavesdropping in the quantum channel. For instance, if an
eavesdropper tries to intercept the QSS scheme by measuring the
state of the particle intended for a legitimate participant using
randomly the $x$ or the $y$ basis, the error rate will be as high
as 25\%, just like the BB84 QKD case.  If the error rate is high,
then Alice concludes that there is eavesdropping and the QSS
session is dropped. If the error rate is low, then the QSS session
is concluded safe and after quantum error correction and privacy
amplification, a final secret sharing bit string is produced. The
$n-1$ participants can determine the shared secret bit using the
QSS rules given in Eqns. (\ref{rule1},\ref{rule2}) for each valid
of transmission.

We have reformulated the HBB QSS protocol in a concise
mathematical form and generalized it into arbitrary multi-parties.
In these rules, the secret key can be simply calculated using the
parity of the measurement outcomes of the participating parties
together with the number of $y$-basis used in the process.
 However only half of the GHZ-states can be used for quantum
secret sharing.

\section{Asymptotic 100\% Efficient QSS Schemes}
\label{s3}

By 100\% efficient, we mean that all the GHZ-state particles used
in a QSS scheme can be used for sharing the secret information as
compared with the original HBB-QSS scheme where half of the
GHZ-states have to be discarded,  because half of the time the
participants may choose an odd number of $\sigma_y$-basis. Here we
propose two efficient QSS schemes using techniques that were
originally used for QKD to increase the efficiency. Full
efficiency can be obtained if the participants  can always choose
the right combination of measuring-basis so that there are always
even number of participants choosing the
$\sigma_y$-measuring-basis. This can be achieved in two different
ways. One is to use the method proposed by Lo, Chau and Ardehali
for QKD\cite{chaulo}. In this scheme the efficiency of the BB84
QKD scheme is asymptotically 100\%. The other one is the one based
on the method proposed by Hwang, Koh and Han\cite{hkh}. They have
discovered that the efficiency of BB84 QKD scheme can be increased
to 100\% by letting Alice and Bob to choose identical
measuring-basis according to a common secret key repeatedly, say
with a 1000 bit control key. For instance a 0 in the control key
means Alice and Bob use the horizontal/vertical measuring basis
and a 1 in the control key directs them to use the
diagonal-antidiagonal measuring basis. These schemes  have several
advantages. First the efficiency is increased to 100\%
asymptotically. Secondly the public announcement of
measuring-basis can be omitted or almost omitted and this saves a
lot of storage space, classical communication and the comparison
computation time. These techniques can be generalized with some
modification for use in QSS. In the following, we present the
results in details.

\subsection{The Favored-Measuring-Basis Efficient QSS scheme}

We call the efficient QSS scheme based on the Lo-Chau-Ardehali
technique as the favored-measuring-basis efficient QSS scheme. It
is noticed that if all the participants in a QSS round choose the
$\sigma_x$-basis, it is a valid QSS round, and  the GHZ-state in
(\ref{ghz3}) can be written in the $\sigma_x$-basis as
\begin{eqnarray}
|\psi\ket_{\rm GHZ}&&=\sqrt{1\over
2^{n+1}}\left(\prod_{i=1}^n(|0\ket+|1\ket)-\prod_{i=1}^n(|0\ket-|1\ket)\right)\nonumber\\
&&=\sqrt{1\over 2^{n-1}} \sum_{i_1i_2\cdots
i_n\;\;}'|i_1\;i_2\;\cdots\;i_n\ket,\label{amb}
\end{eqnarray}
where the prime over the sum means a restricted sum for those
running indices satisfying
\begin{eqnarray}
i_1\oplus i_2 \oplus\cdots\oplus i_n=0. \end{eqnarray} Terms like
$|10\cdots 0\ket$ are absent from the GHZ-state expression because
a part from the second product cancels with that from the first
product term in Eq. (\ref{amb}). Hence a high-efficiency QSS
scheme based on the Lo-Chau-Ardehali technique\cite{chaulo} can be
designed as follows: 1)Alice prepares a sequence of $n$-particle
GHZ-state in state (\ref{ghz3}); 2) For each GHZ-state, Alice
keeps one particle at her own site and sends the rest $n-1$
particles to other participants, each particle to a participant;
3) Each participant chooses with a large probability to measure
his/her particle in the $\sigma_x$-basis, and with a small
probability to measure in the $\sigma_y$-basis. They records the
basis they use and the outcome of the measurement for each
particle; 4) After a large number of GHZ-state particles have been
distributed and measured, they publish their measuring basis for
each GHZ-state; 5) For those rounds of communication where at
least one of the participants chooses the $\sigma_y$-basis, all
the participants publish also the outcomes of their measurements.
 In approximate half of these events, an even number of
participants choose the $\sigma_y$-basis, and the outcomes of the
measurements of all the participants are correlated, and they will
be used to check eavesdropping. We can modify the refined data
analysis method proposed in Ref.\cite{chaulo} to catch Eve. In the
refined data analysis, one only checks those cases that an even
number of participants choose the $\sigma_y$-basis(excluding the
case when no participants choose the $\sigma_y$-basis). Eve's
interception will cause significant errors. Eve needs to intercept
all the $n-1$ particles sent by Alice to the other $n-1$
participants. Suppose Eve always uses the $\sigma_x$-basis to
intercept, for those events that two participants choose the
$\sigma_y$-basis, Eve will introduce an error rate as high as
50\%. If we just look at the events where two participants choose
the $\sigma_y$-basis, this case can be seen as an variant of the
efficient QKD scheme between these two participants where they use
the $\sigma_x$-basis most of the time and the $\sigma_y$-basis
only a small number of times in Ref.\cite{chaulo}.  By examining
the error rate, the participants can determine whether the QSS
communication is secure. For noiseless channels and ideal photon
sources, if no errors exist one can conclude the QSS operations
safe. If there are errors then one concludes that the QSS
operations are insecure and discards the result. For noisy
channels and imperfect photon sources, one has to use quantum
error correction and privacy amplification method to get secure
shared secret information. A more rigorous security analysis for
this scheme, and the details of the post-processing is needed, and
this work in under way. We will not touch this issue in this
paper.

\subsection{The Measuring-Basis-Encrypted Efficient QSS scheme}

We call the efficient QSS scheme based on the  Hwang-Koh-Han QKD
technique as the measuring-basis-encrypted QSS scheme, because the
measuring-basis of the participants are controlled by a secret key
and this information is encrypted. In the Hwang-Koh-Han QKD
scheme, the measuring basis of Alice and Bob in a QKD process is
synchronized by a control key. Different from QKD where Alice and
Bob use the same secret key  to synchronize their measuring-basis,
we need $n$-control keys to control the valid choices of
measuring-basis for the $n$ participants. Furthermore, the control
key sequence is different for different participant. In table
{\ref{t1}, we give an example of control keys for a three party
QSS scheme. Here only the first 10 bits of the control keys are
shown. In practice, the control keys are about 1000 bits long.

The essential part is to generate a control key for each
participant so that the set of measuring-basis in a QSS
transmission always has an even number of $\sigma_y$-basis.  Now
we introduce a method for establishing the control key sequences
for each party on-site using the original HBB QSS scheme. First we
run the HBB QSS scheme in its original form, that is, all parties
choose their measuring-basis randomly. They record their results
and also the measuring-basis information. They then send the
measuring-basis information to Alice, but the measured results are
kept secret. Upon receiving the measuring-basis information from
all parties, Alice can decide which GHZ-multiplets are valid QSS
operation, that is, she knows that in these operations there are
an even number of parties having chosen the $\sigma_y$-basis. She
then tell all the $n-1$ parties to retain the results in these
rounds. Then each of the party will have a sequence of random
numbers which is known only to himself/herself. It is noted that
each party's control key is different from others. These numbers
are used  to determine each party's measuring-basis choice. Except
Alice, each party will use the $\sigma_x(\sigma_y)$-basis, if the
bit value in her/his sequence is a 0(1). Alice's control sequence
is slightly different from the others in the following way: if the
number of $\sigma_y$-basis in the measurement is $4k$, she simply
choose $\sigma_x$($\sigma_y$)-basis if her measured result is a
0(1), and if the number of $\sigma_y$-basis measurement is
$2(2k+1)$, then she will choose $\sigma_y$($\sigma_x$)-basis if
her measured result is a 0(1). This is because when the number of
$\sigma_y$-basis is $4k$, the measured result's parity is even,
and it is odd when it is $2(2k+1)$. As in the QKD case, this
control key can be used repeatedly. The control sequence needs not
be long, a few hundreds of bit, the order of a thousand  is
sufficient.

The on-site generation of the control keys can be spared if the
participating parties keep part of the random numbers left over
from a previous QSS operation.

The security of the QSS has been discussed in Ref.\cite{hbb}, and
the discussion there also applies here. The security of the
repeated use of a control sequence is discussed in
Ref.\cite{hkh,hkhproof}, and they can be adapted here with some
minor modification. The QSS scheme can be viewed as a two 'party'
quantum key distribution scheme if one views the $n-1$ parties as
whole as a single participant. These $n-1$ participants as a whole
share a common secret key with Alice. However inside these $n-1$
parties, they have to act collectively to work out the secret key
of Alice. Any eavesdropping will cause significant errors to the
random key. Similarly, if one of the party is dishonest,
significant error will occur. For instance if Eve uses randomly
the $\sigma_x$-basis and the $\sigma_y$-basis to measure $n-1$
particles Alice sends to the $n-1$ participants, then Eve will
have $(1/2)^{n-1}$ probability to choose the right
measuring-basis. For those that Eve has chosen the wrong
measuring-basis, there is 50\% of probability to make error in the
parity of the string, which is the shared secret information.
Hence the error rate introduced by Eve is \begin{eqnarray}
e=\left(1-(1/2)^{n-1}\right) 1/2. \end{eqnarray} For $n=3$, this
amounts to 3/8=37.5\%.  As the number of participants increase,
the error rate approaches 50\%.

\section{summary}
\label{s4} We have generalized the HBB QSS scheme into arbitrary
number of parties, and given explicit expressions for the shared
secret information in terms of the parity of strings formed by the
measured results of the $n-1$ participants. By generalizing the
Lo-Chau-Ardehali QKD scheme\cite{chaulo} and the Hwang-Koh-Han QKD
scheme\cite{hkh},  we have developed two efficient QSS schemes:
the favored-measuring-basis scheme and the
measuring-basis-encrypted QSS schemes.  The efficiency of these
QSS schemes are asymptotically 100\%.
 We have also qualitatively showed the security of the QSS
scheme. It remains to be shown the security of these QSS schemes
in noisy channels and with imperfect single photon sources,  in a
way similar to what have been done for the security of
QKD\cite{lo1,mayers,shor}. Work is under way and the result will
be published elsewhere.

This work is supported the National Fundamental Research Program
Grant No. 001CB309308, China National Natural Science Foundation
Grant Nos. 60073009, 10325521, the Hang-Tian Science Fund, and the
SRFDP program of Education Ministry of China.

\begin{table}
\begin{center}
\caption{ An example of valid control keys for an three-party
measuring-basis-encrypted QSS scheme}\label{t1}
\begin{tabular}{ccccccccccc}\\ \hline\hline
Round No. & 1 & 2 & 3 & 4 & 5 & 6 & 7 & 8 & 9 & 10\\ \hline
Alice
& $x$ & $y$ & $x$ & $y$ & $x$ & $x$ & $y$ & $x$ & $x$ &
$y$ \\
Bob       & $x$ & $x$ & $y$ & $y$ & $x$ & $x$ & $y$ & $y$ & $x$ & $x$ \\
Charlie   & $x$ & $y$ & $y$ & $x$ & $x$ & $x$ & $x$ & $y$ & $x$ & $y$\\
\hline
\end{tabular}
\end{center}
\end{table}
\end{document}